\documentclass[11pt]{iopart}
\usepackage{iopams} 
\usepackage[colorlinks=true,citecolor=blue,urlcolor=blue,linkcolor=blue]{hyperref}
\usepackage{graphicx}
\usepackage{tabu}
\usepackage{makecell}  
\usepackage{cite}
\usepackage{subfig} 

\expandafter\let\csname equation*\endcsname\relax
\expandafter\let\csname endequation*\endcsname\relax
\makeatletter
\newcommand*{\defeq}{\mathrel{\rlap{%
                     \raisebox{0.3ex}{$\m@th\cdot$}}%
                     \raisebox{-0.3ex}{$\m@th\cdot$}}%
                     =}
\makeatother
\usepackage{amsmath}
\newcommand{\ket}[1]{|#1\rangle}
\newcommand{\bra}[1]{\langle #1|}
\newcommand{\abs}[1]{| #1 |}
\newcommand{\braket}[2]{ \langle #1 |#2\rangle}
\newcommand{\ketbra}[2]{|#1\rangle\langle #2|}
\newcommand\Tstrut{\rule{0pt}{2.6ex}}        
\newcommand\Bstrut{\rule[-0.9ex]{0pt}{0pt}}  
\newcommand{\TBstrut}{\Tstrut\Bstrut}        

\makeatletter
\newcommand{\thickhline}{%
    \noalign {\ifnum 0=`}\fi \hrule height 1pt
    \futurelet \reserved@a \@xhline}

\begin{document}
\title[The walker speaks its graph]{The walker speaks its graph: 
global and nearly-local probing of 
the tunnelling amplitude in continuous-time quantum walks}
\author{Luigi Seveso\footnote{luigi.seveso@unimi.it}, Claudia Benedetti\footnote{claudia.benedetti@unimi.it} \& Matteo G. A. Paris\footnote{matteo.paris@fisica.unimi.it}}
\address{Quantum Technology Lab, Dipartimento di Fisica Aldo Pontremoli, Universit\`a degli Studi di Milano, I-20133 Milano, Italy}
\begin{abstract}
We address continuous-time quantum walks on graphs, and 
discuss whether and how quantum-limited measurements on 
the walker may extract information on the tunnelling amplitude 
between the nodes of the graphs. For a few remarkable families 
of graphs, we evaluate the ultimate quantum bound to precision, 
i.e. we compute the quantum Fisher information (QFI), 
and assess the performances of incomplete
measurements, i.e. measurements performed on a subset 
of the graph's nodes. We also optimize the QFI over the 
initial preparation of the walker and find the optimal 
measurement achieving the ultimate precision in each case. 
As the topology of the graph is changed, a non-trivial 
interplay between the connectivity and the achievable 
precision is uncovered. 
\end{abstract}
\submitto{\JPA}
\section{Introduction}
Continuous-time quantum walks (CTQWs) generalize classical random walks to the quantum domain, by modeling the  propagation of a quantum particle over a discrete space continuously in time \cite{fahri98,kempe,venegas}. Recently, quantum walks have found powerful applications in different quantum computing tasks, e.g.~they provide a model for universal quantum computation \cite{childs09} and represent the building blocks of several quantum algorithms \cite{childs04,kendon06,ambainis07,farhi08,coppersmith10}. Moreover, quantum walks allow to study the transport properties of excitations on networks \cite{mulken11,alvir16,tamas16} and are employed to simulate the dynamics of some biological systems \cite{mohseni08,hoyer10}. As far as experimental implementations are concerned, quantum walks have been realized on a number of physical architectures, from trapped ions systems \cite{preiss2015strongly} to photonic platforms \cite{peruzzo2010quantum}. The ubiquity of quantum walks in quantum information is due to the fact that the dynamics of any physical system whose Hilbert space is comprised of (or truncated to) a finite number of discrete states can always be mapped into a quantum walk on a graph \cite{hines07}, i.e.~a discrete mathematical structure made up of a set of nodes, or vertices, connected by edges. Each node of the graph coincides with a system's state and it is put in one-to-one correspondence with one of the walker's position eigenstates, while each edge is assigned a positive real weight, which is equal to the transition amplitude between the states corresponding to the two endpoints. 
\par
In this paper, we focus on the following theoretical task: to fully reconstruct the Hamiltonian of a quantum walker from a series of measurements on it; or, equivalently, to statistically infer the numerical values of the weights of the graph corresponding to a given quantum walk implementation \cite{bugart09,hillery12,tama16,nokkala18}.
It is important to notice that our approach is quite different from most lines of research, where the complete knowledge of the graph parameters are used to infer the behavior of the quantum particle. 
For instance, quantum dots can be used to implement quantum walks on cycle graphs 
\cite{ito16}, 
  and can be addressed to study the properties of the quantum particle by having full
 control of the graph parameters \cite{mlknikov16,zwolak18}.
On the contrary, here we want to extract information about the graph parameters by observing the 
dynamics of a quantum walker.

We tackle the problem in a rigorous way from a quantum parameter estimation perspective \cite{paris09}, under the simplifying assumption that all weights are set equal to the same constant, which determines the tunnelling amplitude of the walker, and that the underlying graph belongs to one of a few remarkable classes introduced below. By performing repeated measurements on identically prepared states of the walker and collecting the resulting outcomes, one can estimate the tunnelling amplitude via a suitable estimator. In particular, the quantum Fisher information (QFI) quantifies the maximum amount of information that can be extracted by any estimation protocol, optimized over all possible quantum measurements. In the following, we compute the QFI for different graph topologies, which leads to uncover a rich phenomenology, as a function, e.g., of the graph's connectivity, the number of nodes and the interrogation time. The ultimate QFI limit is then compared with the performance of a few specific measurements that are assumed to be available to the experimentalist, focusing in particular on measurements that require access only to a subset of the graph's nodes.
\par
The rest of the paper is organized as follows. In Section~\ref{sec:qw}, we define the concept of continuous-time quantum walk and introduce a number of special graph families. In Section~\ref{sec:qet}, we review the basic tools of quantum parameter estimation theory. Section~\ref{sec:res} is devoted to the analysis of the best achievable precision, which is then compared with the performance of a few realistic measurements. Finally, in Section~\ref{sec:conc} we draw our conclusions.
\section{Continuous-time quantum walks on graphs}\label{sec:qw}
A CTQW takes place in the Hilbert space $\mathcal{H}=\mathbb{C}^n$ spanned by the position basis of the walker, which is denoted by $\{\ket{k}\}_{k=1}^n$, with $\braket{k}{j}=\delta_{kj}$. Each vector $\ket{k}$ represents a state of the walker localized at position $k$. Formally, a CTQW corresponds to the Hamiltonian
\begin{equation}
H=-\sum_{(j,k)\in E} \gamma_{jk}(\ketbra{j}{k}+\ketbra{k}{j})+\sum_{j\in V}\epsilon_j\ketbra{j}{j}\;,
\label{h1}
\end{equation}
where the coefficients $\gamma_{jk}$ represent the tunneling amplitudes between adjacent positions and $\epsilon_j$ the local energies at each site. The particle moves on a weighted graph $G(V,E)$, where $V=\{k\}$ is the set of nodes or vertices, with cardinality $n = |V|$, while $E$ is the set of edges.  The edge between nodes $j$ and $k$ is associated with the weight $\gamma_{jk}$, whereas the nodes have weights $\epsilon_j$. Notice that a specific conventional labelling for the nodes of the graph has been assumed. In the following, we restrict ourselves to the following particular case: we take all weights to be equal, i.e. $\gamma_{jk}=\gamma\in \mathbb{R}^+$, $\forall\, j,k=1,\dots n$, and $\epsilon_j=\epsilon d_j$, with $d_j$ the degree of vertex $j$. If in addition we set $\epsilon=1$, i.e. we measure all quantities in units of $\epsilon$, then the Hamiltonian of Eq.~\eqref{h1} can be rewritten as:
\begin{equation}
H_G=D -\gamma A\;,
\label{hami}
\end{equation}
where $A$, is the  adjacency matrix of the graph $G$, i.e.~a square $n\times n$ matrix whose elements are $A_{jk}=1$ if nodes $j$ and $k$ are connected and zero otherwise, and $D$ is the degree matrix, i.e.~a diagonal matrix  whose entries  $D_{jj}=\sum_k A_{jk}$ are the vertex degrees. The tunnelling amplitude  $\gamma$ is therefore the only relevant parameter to be estimated: its value determines the amplitude of the transition between different states (nodes) and depends on the particular quantum walk implementation. 
\par
\begin{figure}
\begin{center}
\begin{tabular}{lllllllllll}
\subfloat[]{\includegraphics[width = 0.75in]{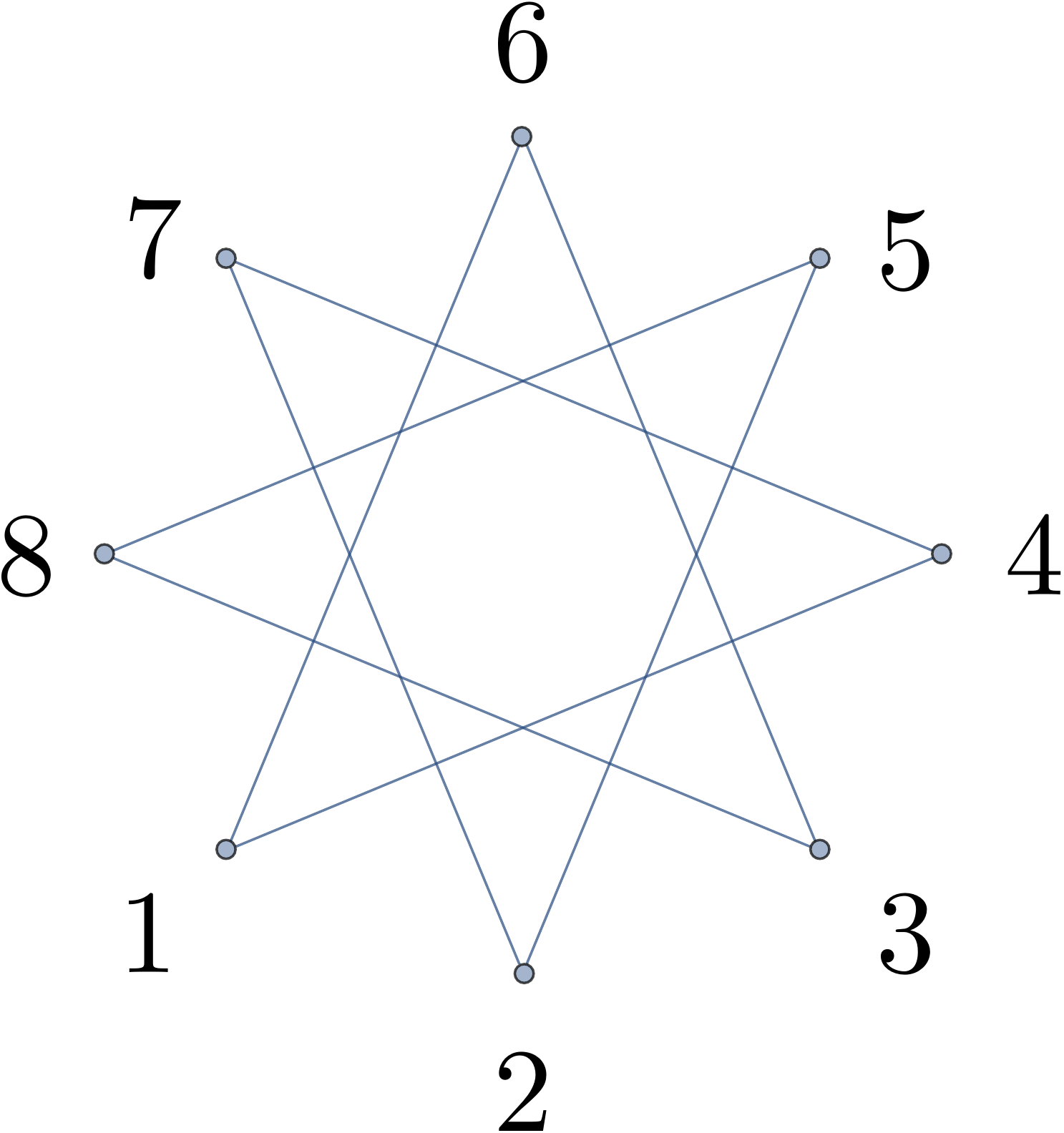}} &
 &
\subfloat[]{\includegraphics[width = 0.75in]{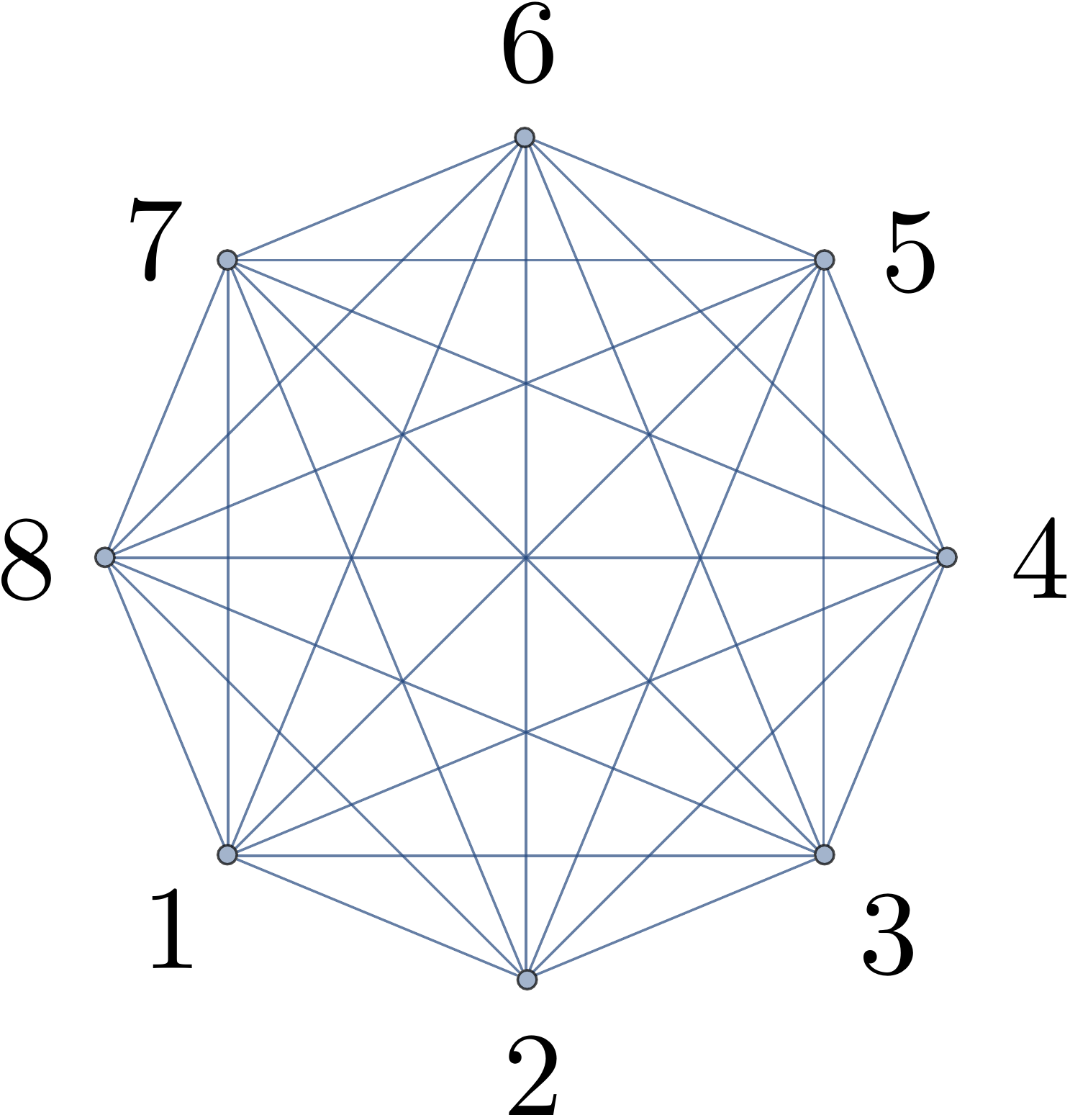}} &
 &
\subfloat[]{\includegraphics[width = 0.75in]{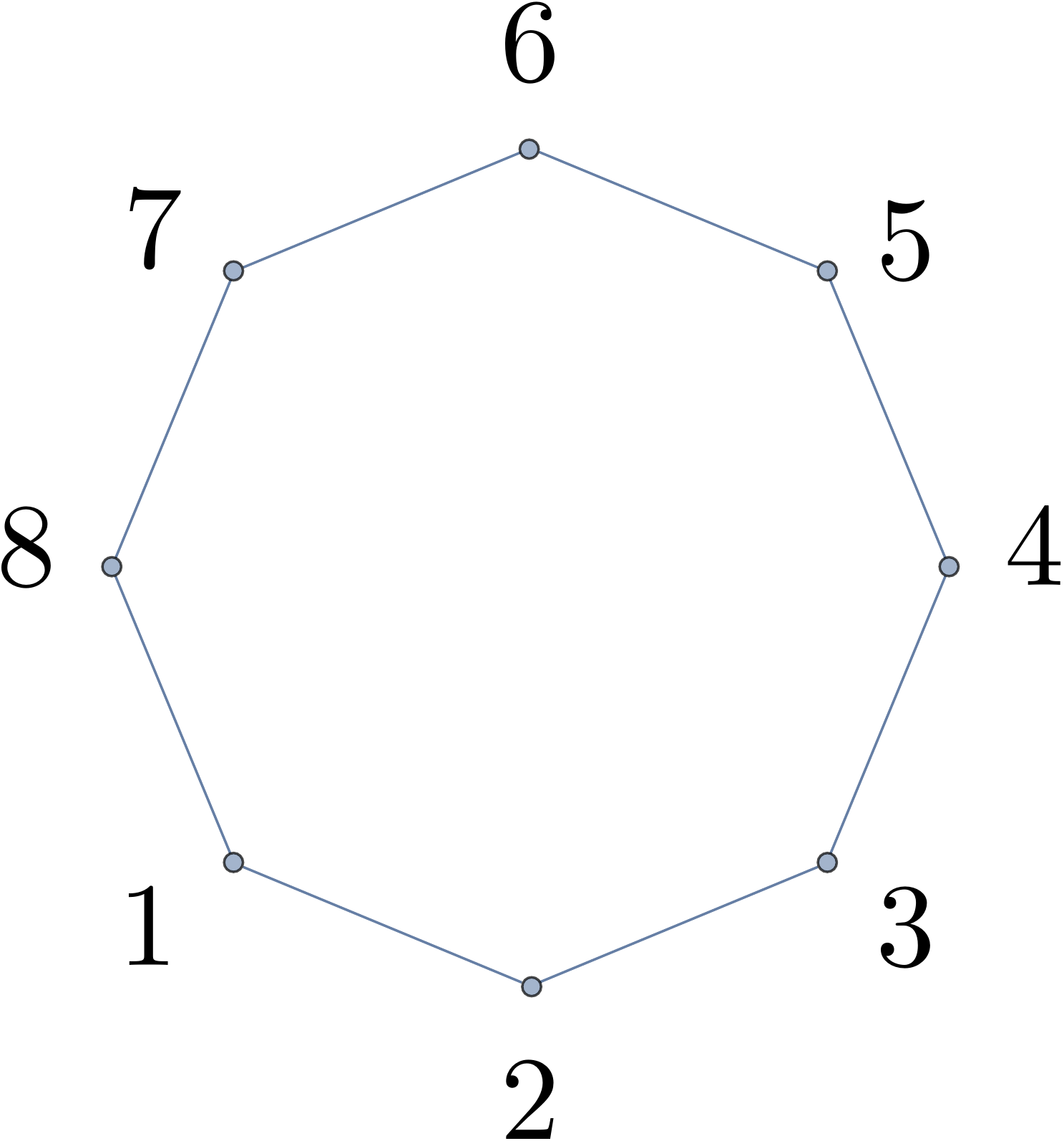}} &
 &
\subfloat[]{\includegraphics[width = 0.75in]{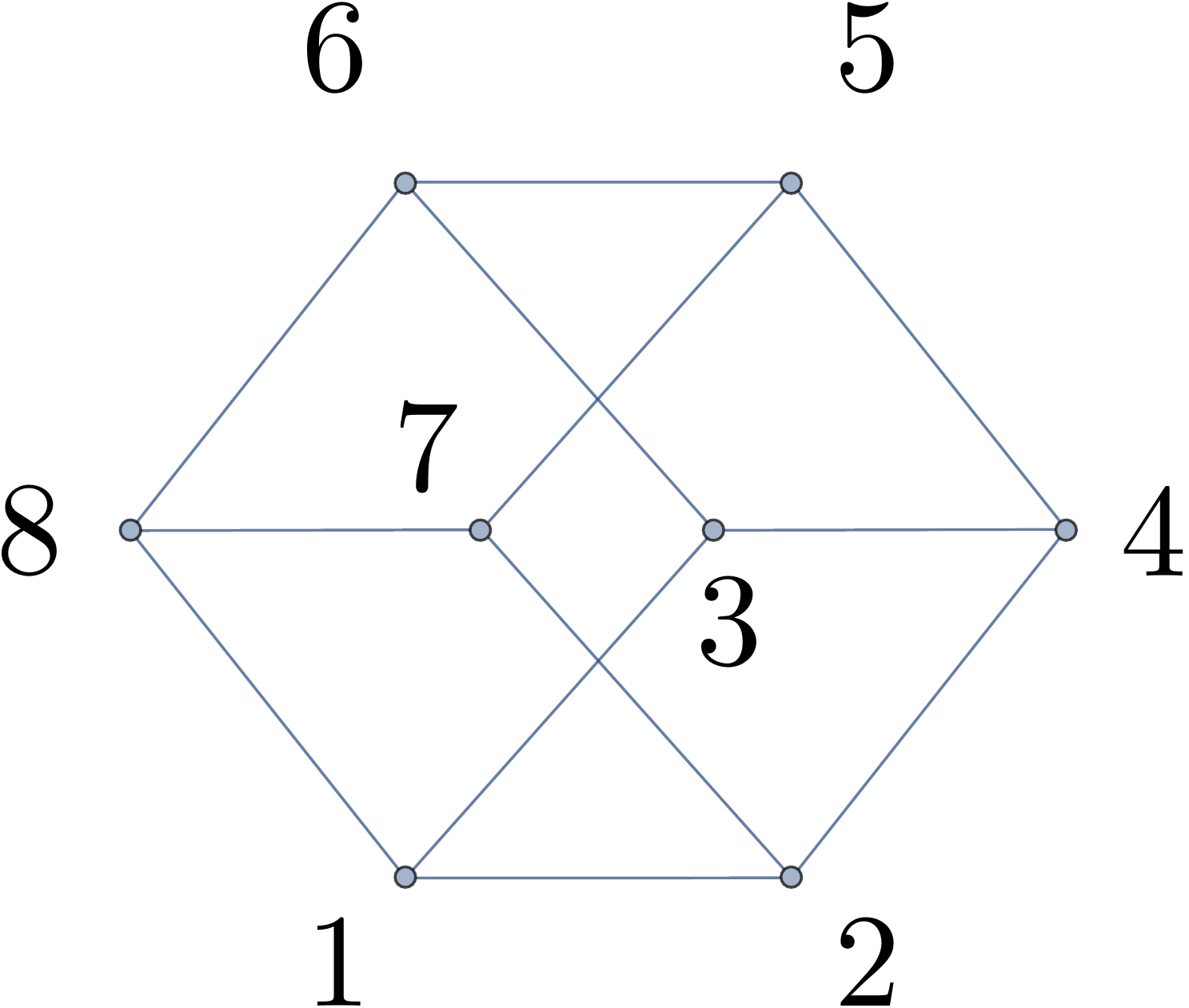}} &
 &
\subfloat[]{\includegraphics[width = 0.75in]{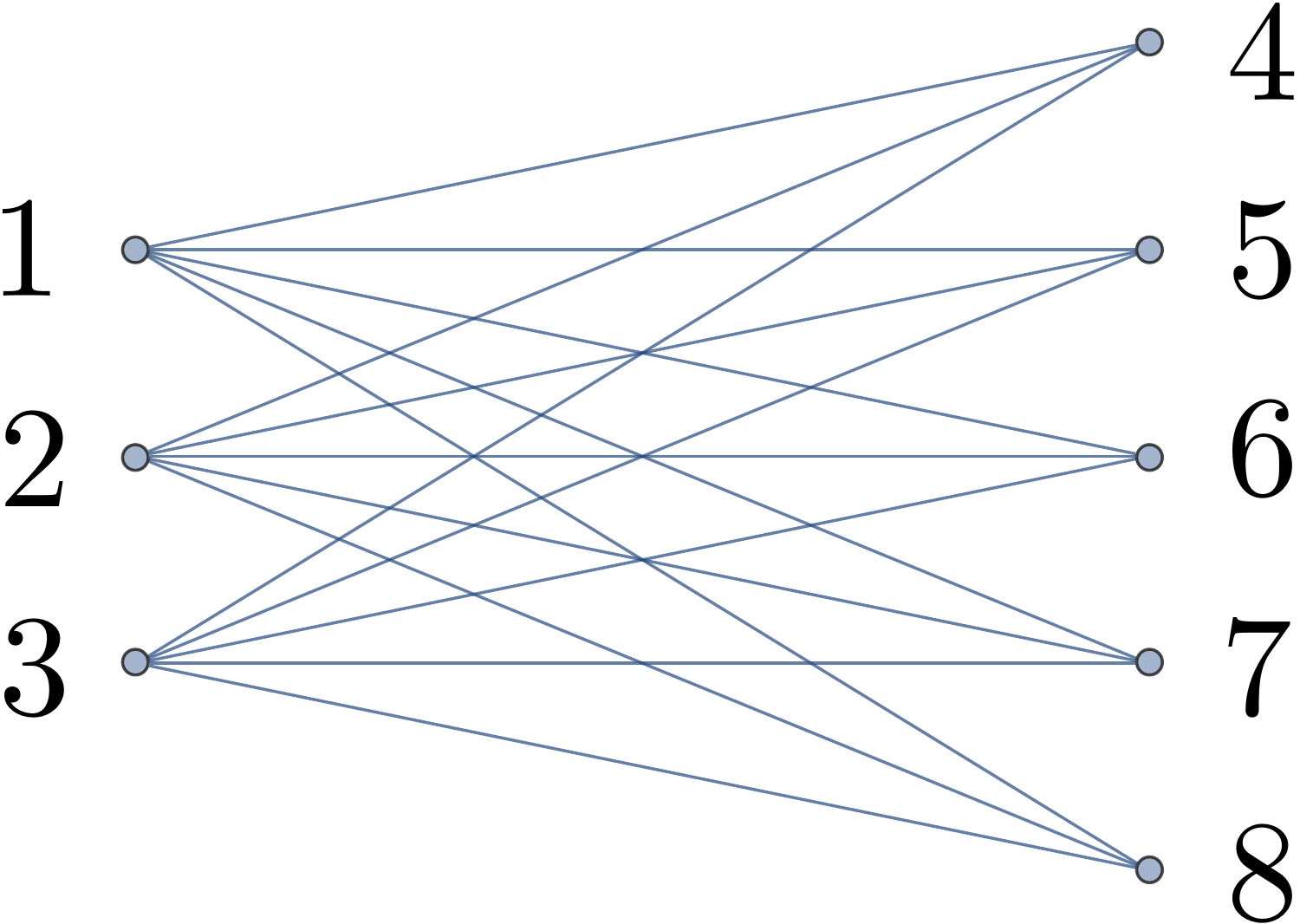}} &
 &
\subfloat[]{\includegraphics[width = 0.75in]{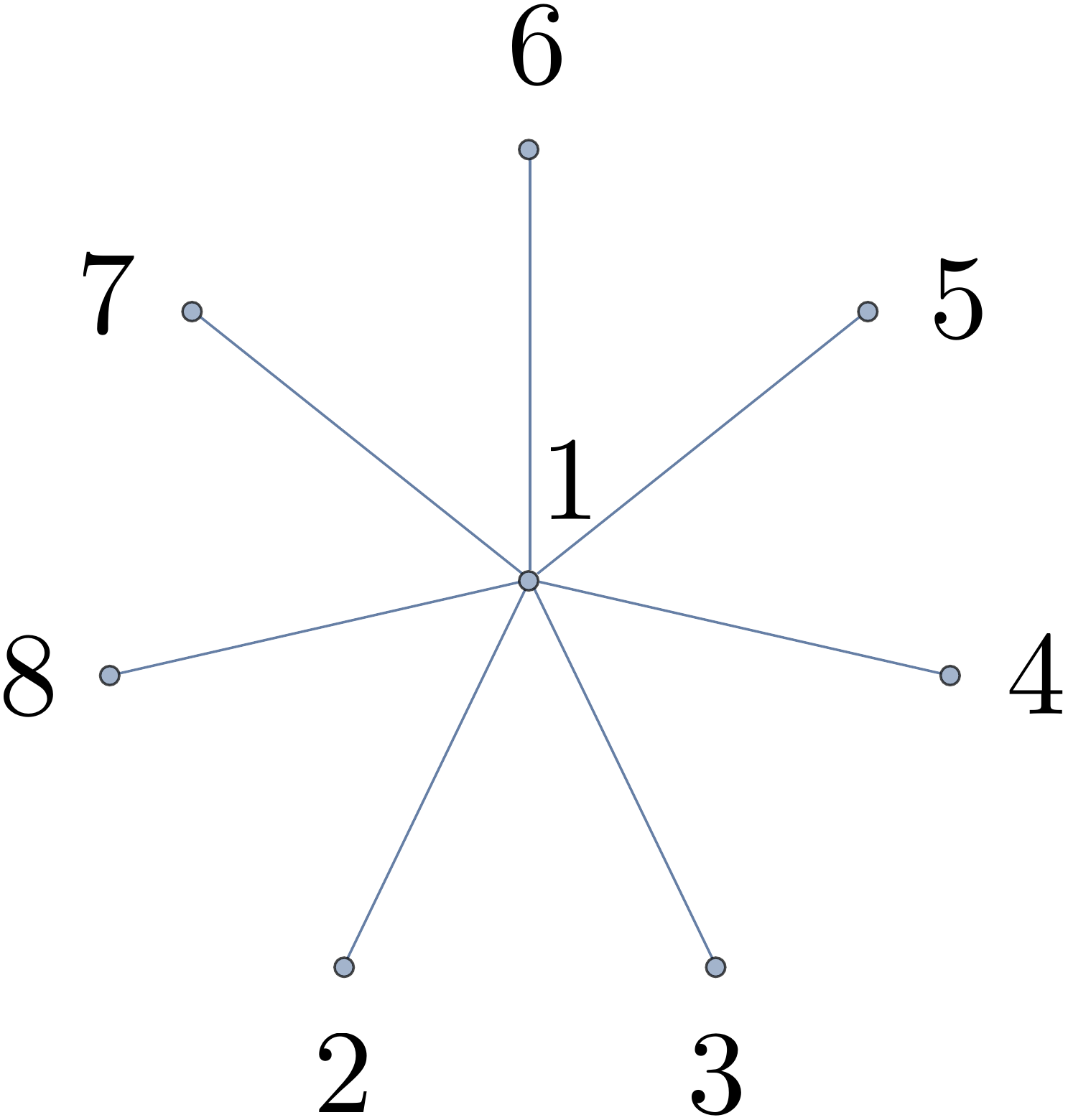}}
\end{tabular}
\caption{Families of graphs considered in the main text, with a conventional choice of labelling: \textbf{(a)} circulant graph, \textbf{(b)} complete graph, \textbf{(c)} cycle graph, \textbf{(d)} hypercube graph, \textbf{(e)} complete bipartite graph, \textbf{(f)} star graph. \label{picGraph}}
\end{center}
\end{figure}

We will consider in detail several remarkable families of simple graphs, i.e.~graphs that are undirected, with no loops or multiple edges between any two vertices. The state of the walker at time $t$  is given by $\ket{\psi_t}=U_t \ket{\psi_0}$, where $\ket{\psi_0}$ is the initial state and $U_t=\exp(-i H_G t)$ is the evolution operator.  If $\xi_j$ and $\ket{\xi_j}$ (with $j\in \{0\dots,n-1\}$) are respectively the eigenvalues and the eigenstates of the Hamiltonian $H_G$, and the particle is initially prepared in an arbitrary superposition $\ket{\psi_0}=\sum_{j=0}^{n-1}\alpha_j\ket{\xi_j}$, the  state of the walker after an interrogation time $t$ can be written as $\ket{\psi_t}=\sum_{j=0}^{n-1}\alpha_j e^{-i \xi_j t}\ket{\xi_j}$. Thus, knowledge of the eigenvalues and eigenvectors of $H_G$ is sufficient to completely characterize any quantum walk on $G$. For this reason, we are now going to review the spectral properties of a few special families of graphs on which we will focus our attention later on. More precise definitions and derivations can be found in \ref*{appe} (see also Fig.~\ref{picGraph}).
\subsection{Circulant graphs: complete and cycle graphs}
A circulant graph $O_n$ is a simple graph having the following property: there exists a relabelling of its vertices which is (1) an isomorphism (i.e., any two vertices of the resulting graph are connected by an edge iff they are so connected before the relabelling) and (2) is a cyclic permutation of the vertices (i.e., the relabelling permutes a subset of the vertices in a cyclic fashion, while leaving fixed the remaining ones). 

Two special cases of circulant graphs are complete graphs and cycle graphs. A \emph{complete} graph $K_n$ is a graph where each vertex is adjacent to any other vertex. It has $n$ vertices, $n(n-1)/2$ edges and it is regular with degree $n-1$. In  the case of a weighted complete graph with all couplings equals, the eigenvalues of the Hamiltonian $H_{K_n}$ as in Eq.~\eqref{hami} are:
\begin{equation}\label{eigvaluesComplete}
	\xi_j  =
	\begin{cases}
	 (n-1)(1 - \gamma)  \qquad\; \text{if $j=0$}\\
	 (n-1) + \gamma \,\quad\qquad\;\, \text{if $j\neq 0$}
	\end{cases}\;.
	\end{equation}
The corresponding eigenvectors are
\begin{equation}\label{eigvectcomp}
\ket{\xi_j} = \frac{1}{\sqrt n}\sum_{k=1}^{n} e^{\frac{2\pi i j (k-1)}{n}}\,\ket{k}\;,
\end{equation}
where $\ket{k}$ is the position vector corresponding to the $k^{\text{th}}$ vertex.

A \emph{cycle} graph $C_n$ is a chain of $n$ vertices (and as many edges). It is circulant and regular with degree $2$. By specializing to this case Eqs.~\eqref{geneigval} and \eqref{eigvec}, found in \ref*{appe}, the eigenvalues of $C_n$ are found to be
\begin{equation}
\xi_j= 2 - 2\gamma\cos\left(\frac{2\pi  j}{n}\right)\;,
\end{equation}
while the eigenvectors are the same as in Eq.~\eqref{eigvectcomp}.
%
\subsection{Hypercube graphs}
A \emph{hypercube} graph $Y_d$ is a regular graph formed from the vertices and edges of the 
$d$-dimensional hypercube. It has $n=2^d$ vertices and $d\cdot 2^{d-1}$ edges; besides, each 
vertex has the same degree $d$. It can also be  thought of as the $d$-fold Cartesian product of 
the complete graph $K_2$.  The eigenvalues of the Hamiltonian $H_{Y_d}$ for a hypercube graph in dimension $d$ evaluate to
\begin{equation}\label{eigValuesHyper}
\xi_j = d - \gamma(d-2j)\;,
\end{equation}
while the eigenvectors are computed recursively in the position eigenbasis as the columns of the Hadamard matrices $B_d$, $d\in \mathbb N$, defined as follows:
\begin{equation}\label{FRec}
B_1 = \frac{1}{\sqrt 2} \begin{pmatrix}
1 & 1 \\
1 & -1
\end{pmatrix}\;,\qquad \quad
B_d = \frac{1}{\sqrt 2}\begin{pmatrix}
B_{d-1} & B_{d-1} \\
B_{d-1} & - B_{d-1}
\end{pmatrix}\;.
\end{equation}
%
\subsection{Complete bipartite graphs}
A \emph{complete bipartite} graph $K_{p,q}$ is such that its vertex set $V$ can be 
split into two subsets, with the following property: all vertex of the first set are 
adjacent to every other vertex of the second, but no two vertices in either sets are 
adjacent among themselves. It has $n=p+q$ vertices, with $\{1,\dots, p\}$ and 
$\{p+1,\dots, p+q\}$ denoting the two partitions. A special case of a complete 
bipartite graph is the star graph $S_n = K_{1,n-1}$, corresponding to setting 
$p=1$ and $q=n-1$. The degenerate spectrum of $H_{K_{p,q}}$ is comprised of the eigenvalues $\{\xi_1,\xi_2,\xi_+,\xi_-\}$, where
\begin{equation}\label{xiDet}
 \xi_1=q\;,\qquad \xi_2=p\;,\qquad  \xi_{\pm} = \frac{(p+q)\pm\sqrt{\Delta_{p,q}}}{2}\;,
\end{equation}
and we defined $\Delta_{p,q} \defeq (p-q)^2 +4qp\,\gamma^2$. The corresponding eigenstates, written in the position eigenbasis, are
\begin{equation}\label{eigvecstot}
\ket{\xi_1^{(\kappa)}} = (\boldsymbol a^{(\kappa)},\,\boldsymbol 0_q)^t\;,\qquad
\ket{\xi_2^{(\kappa')}} = (\boldsymbol 0_p,\,\boldsymbol b^{(\kappa')})^t\;,\qquad
\ket{\xi_\pm} = \eta_\pm\left(\frac{p  - \xi_\pm}{\gamma p}\,\boldsymbol 1_p,\, \boldsymbol 1_q\right)^t\;, 
\end{equation}
where $\eta_\pm \defeq 1/\sqrt{q\left[1+(p-\xi_\pm)/(q- \xi_\pm)\right]}$, $\boldsymbol 0_q$ is a $q$-vector made up of all zeros, $\boldsymbol 1_q$ is a  $q$-vector made up of all ones, and $\kappa,\, \kappa'$ are two degeneracy indexes (see \ref*{appe} for more details).
\section{Quantum estimation theory}\label{sec:qet}
The typical setting in quantum parameter estimation theory is given by a parametric family of quantum states $\{\rho_{\gamma}\}$, smoothly depending on a real parameter $\gamma$. It is assumed that there exists a true value $\gamma^*$ such that $\rho_{\gamma^*}$ describes the state of the system available to the experimentalist. The main task is to infer the true value  from the outcomes of repeated measurements on the system. In fact, if there is no Hermitian operator whose eigenvalues correspond to the possible values of $\gamma$, the parameter cannot be measured directly, and the only possibility is to rely on indirect measurements in order to infer it. More precisely, by preparing $N$ identical copies of the system and repeating $N$ times a measurement $\mathcal M$ with sample space $\mathcal X$, one obtains a  sample $\Omega=\{x_1,x_2\dots,x_N\}\in \mathcal X^{\times N}$, which is then processed via a local quantum estimator. An estimator is a function $\hat{\gamma}: \Omega\rightarrow \mathbb{R}$ that maps the sample $\Omega$ into an estimate of the parameter. In any quantum estimation procedure, the aim is to optimize over the choice of the measurement in order to maximize the information about the parameter, i.e.~to minimize the mean square error of the estimator $\hat{\gamma}$. Because of the inherent stochasticity of quantum measurements, there are strict limitations on the achievable precision. Indeed, the variance of any unbiased estimator is bounded by the Cram\'er-Rao bound (CRB) as follows:
\begin{equation}
\text{Var}(\hat{\gamma})\geq\frac{1}{N\cdot \mathcal F( \mathcal{M},\gamma)}\;,
\end{equation}
where $\mathcal F( \mathcal{M},\gamma)$ is the Fisher information (FI) associated with the measurement $ \mathcal M$, i.e.,
\begin{equation}
\mathcal F( \mathcal{M},\gamma)=\sum_{x\in\mathcal X}p(x|\gamma)\left[\partial_\gamma \log p(x|\gamma)\right]^2\;,
\label{fishin}
\end{equation}
with $p(x|\gamma)= \tr(\rho_\gamma \Pi_x)$ the conditional probability of obtaining the result $x$ for a measurement $\mathcal M$ having positive-operator valued (POVM) elements $\{\Pi_x\}_{x\in \mathcal X}$, assuming that the value of the parameter is $\gamma$. Minimizing the variance of unbiased estimators is equivalent to maximizing their FI. Therefore, one defines the quantum Fisher Information (QFI) as
\begin{equation}
\mathcal{Q}(\gamma)=\underset{\mathcal M}{\max}\;\mathcal F( \mathcal{M},\gamma)\;,
\end{equation}
which leads to the quantum Cram\'er-Rao bound (QCRB) \cite{helstrom,Fujiwaraa95,petz96,brody98,liu14},
\begin{equation}
\text{Var}(\hat{\gamma}) \geq \frac{1}{N\cdot \mathcal{Q}(\gamma)}.
\label{qcr}
\end{equation}
The QCRB establishes the ultimate limit to precision that is imposed by the quantum mechanical nature of the problem. It can always be saturated, at least in the asymptotic regime $N \to \infty$, by employing an asymptotically efficient estimator and implementing the optimal Braunstein-Caves measurement $\mathcal M_{\text{opt}}$ \cite{braunstein94, braunstein96, nagaoka}, defined as:
\begin{equation}
\mathcal M_{\text{opt}} \defeq \underset{\mathcal M}{\text{arg max}}\;\mathcal F( \mathcal{M},\gamma)\;.
\end{equation} 
Remarkably, for pure states $\ket{\psi_{\gamma}}$, a closed-form expression for the QFI is available \cite{paris09},
\begin{equation}\label{qfi}
\mathcal{Q}(\gamma)=4\big[\braket{\partial_{\gamma}\psi_{\gamma}}{\partial_{\gamma}\psi_{\gamma}}+\big(\braket{\partial_{\gamma}\psi_{\gamma}}{\psi_{\gamma}}\big)^2\big]\;,
\end{equation}
where $\ket{\partial_{\gamma}\psi_{\gamma}}$ denotes the derivative of the statistical model $\ket{\psi_{\gamma}}$ with respect to the parameter $\gamma$. 
\par
In our specific case, the statistical model coincides with the state of the walker at time $t$. As a result, the corresponding QFI still depends on the initial preparation, i.e.~on the coefficients $\alpha_j$. Therefore, the final step is to maximize the QFI over the choice of such coefficients. In this regard, a useful inequality, to which we will often resort in the following, is 
Popoviciu's inequality \cite{popoviciu35}. Given a bounded probability distribution describing a classical random variable $\mathbb Y$, whose minimum and maximum values are denoted by $y$ and $Y$ respectively, Popoviciu's inequality  states that the variance $\text{Var}(\mathbb Y)$ is bounded as follows,
\begin{equation}
\text{Var}(\mathbb Y) \leq \frac{1}{4}(Y-y)^2\;. \label{popineq}
\end{equation}
Equality holds whenever half of the probability distribution is peaked at each of the two values.
\section{Maximum extractable information vs. performance of selected measurements}\label{sec:res}
Given a quantum walk on a graph $G$ with Hamiltonian $H_G$ as  in Eq.~\eqref{hami}, the problem we are going to consider is to quantify how precisely the tunnelling amplitude $\gamma$ can be estimated via measurements on the walker. 
In particular, in this section we compute the maximum amount of information on $\gamma$ that can be extracted via any quantum measurement, i.e.~the quantum Fisher information $\mathcal{Q}(\gamma)$. Because the QCRB is tight, there always exists a quantum measurement whose FI equals the QFI. However, the optimal measurement may be quite exotic, or even depend on the true value of the parameter, so that it is not necessarily available to the experimentalist.  For this reason, we also analyze the performances of a few specific measurements and compare them with the QFI limit.
We focus in particular on position measurements. By definition, a position measurement leads to a projection onto the walker's position basis, i.e.~the position operator is $\hat x = \sum_{i=1}^n j\, \ket{j}\bra{j}$. It is also useful to introduce incomplete position measurements. They correspond to coarse-grainings of a position measurement. Explicitly, an incomplete position measurement of size $m$ is represented by a POVM made up of $m$ rank-1 projectors onto $m$ given position eigenstates, plus the projector onto the orthogonal complement of the subspace spanned by them. Incomplete measurements model a situation where one has experimental access only to a subset of the graph's  nodes.
\par
In the following, we first focus on the QFI, studying in particular how it scales with the number of vertices $n$ and the interrogation time $t$, and maximizing it over the initial preparation of the probe. Then, we analyze the performance of position measurements for different graph families, assuming that the initial preparation coincides with the optimal one maximizing the QFI.
\subsection{Circulant graphs: complete and cycle graphs}
\subsubsection{Complete graphs}
Let us consider the generic {\bf complete graph} $K_n$. The initial preparation is taken to be $\ket{\psi_0} = \sum_{j=0}^{n-1} \alpha_j \ket{\xi_j}$, where $\ket{\xi_0}$ is the ground state and the other energy eigenstates span a degenerate subspace (the corresponding eigenvalues are given in Eq.~\eqref{eigvaluesComplete}). The QFI can be easily evaluated via Eq.~\eqref{qfi}, leading to:
\begin{equation}
 \mathcal Q_{K_n}(\gamma) =  4 n^2 t^2\,\abs{\alpha_0}^2 (1-\abs{\alpha_0}^2)\;.
\end{equation}
Maximizing over the initial preparation, one has that
\begin{equation}\label{limK}
\underset{\ket{\psi_0}}{\text{max}}\;\mathcal Q_{K_n}(\gamma)=n^2 t^2,
\end{equation}
which is obtained when $\ket{\psi_0}$ is equally distributed between the ground state $\ket{\xi_0}$ and the excited energy subspace. Notice that the maximum QFI scales quadratically with both the number of vertices $n$ and the interrogation time $t$. 

We now want to investigate whether a realistic measurement on the walker allows us to attain the QFI limit of Eq.~\eqref{limK}. We thus assume that the state of the walker at time $t=0$ is  $\sum_{j=0}^n \alpha_j \ket{\xi_j}$ with 
$\alpha_0=\alpha_1=1/\sqrt2$ and all other $\alpha_j$ set to zero. After a time $t$, an incomplete position measurement of size $m$ is performed. By definition, its POVM consists of the projectors $\ket{j}\bra{j}$, for $j\in\{1,\dots,m\}$, plus the projector onto their orthogonal complement $\mathbb I_n -  \sum_{j=0}^m \ket{j}\bra{j}$. The corresponding FI is denoted by $\mathcal F_{K_n}^{(m)}(\gamma)$. The efficiency is defined as the ratio of the FI to the QFI, i.e.
\begin{equation}
\eta^{(m)} \defeq \frac{\mathcal F_{K_n}^{(m)}(\gamma)}{\mathcal Q_{K_n}(\gamma)}\,.
\end{equation}
From Eq.~\eqref{fishin}, the Fisher information $\mathcal F_{K_n}^{(m)}(\gamma)$ can be written as
\begin{equation}\label{FICompl}
	   \mathcal F_{K_n}^{(m)}(\gamma) = \sum_{j=1}^m \frac{(\partial_\gamma p_j)^2}{p_j} + \frac{(\partial_\gamma \bar p)^2}{\bar p}\;,
	\end{equation}
	where $p_j$ is the probability of detecting the walker at node $j$, i.e.
	\begin{equation}
	  p_j = \abs{\braket{j}{\psi_t}}^2 = \frac{2}{n}\,\cos^2\left[\frac{\gamma nt}{2} - \frac{\pi(j-1)}{n}\right]\;,
	\end{equation}
	and $\bar p \defeq 1 -  \sum_{j=1}^m p_j$ is the probability that the walker is located outside the subset of the graph's nodes under control by the experimentalist. It is possible to rewrite $\bar p$ as follows,
	\begin{equation}
	  \bar p = \frac{n-m}{n} -  \frac{1}{n} \cos\left[\gamma n t - \frac{(m-1)\pi}{n}\right]\,\frac{\sin\left(\frac{\pi m}{n}\right)}{\sin\left(\frac{\pi}{n}\right)}\;.
	\end{equation}
The proof requires first to transform $\cos^2(x/2)=(1+\cos x)/2$ and then to simplify the sum over the cosine terms via the geometric series identity
\begin{equation}
  \sum_{j=0}^{m-1} e^{i\left(x - \frac{2\pi j}{n}\right)} = e^{ix} \left(\frac{1-e^{-\frac{2\pi i m}{n}}}{1-e^{-\frac{2\pi i }{n}}}\right)\;.
\end{equation}
The final result for $\mathcal F_{K_n}^{(m)} (\gamma)$ is
\begin{equation}\label{FIComplm}
	 \mathcal F_{K_n}^{(m)} (\gamma) = nt^2\left(\!m -  \cos\!\left[ \gamma n t - \scriptstyle \frac{(m-1)\pi}{n}\right] s_{n,m}  
 +\frac{ \sin^2\!\left[\gamma nt -\scriptstyle\frac{(m-1)\pi}{n}\right]s_{n,m}^2}{(n\!-\!m) \!-\!  \cos\!\left[\gamma n t -\scriptstyle \frac{(m-1)\pi}{n}\right] s_{n,m}}\!\right)\;,
\end{equation}
where $s_{n,m}\defeq \frac{\sin\left(\frac{\pi m}{n}\right)}{\sin\left(\frac{\pi}{n}\right)}$. For instance, if $m=1$:
\begin{equation}
   \mathcal F_{K_n}^{(1)}(\gamma) = \frac{2 \sin^2\left(\frac{\gamma n t}{2}\right)}{n - 2 \cos^2\left(\frac{\gamma n t}{2}\right)} n^2 t^2\;.
\end{equation}
The corresponding efficiency, optimized over the interrogation time $t$, is $\text{max}_{\,t}\; \eta^{(1)} = \frac{2}{n}$. That is, a local measurement can at most extract a fraction $2/n$ of the maximum available information. Vice versa, for a complete position measurement, i.e.~$m=n$, one finds from Eq.~\eqref{FIComplm} that $\mathcal F^{(n)}_{K_n}(\gamma) = n^2 t^2$, which is also equal to the QFI: a complete position measurement is optimal. For a generic incomplete measurement with $1 < m < n$, one finds:
\begin{equation}
\underset{t}{\text{max}}\; \eta^{(m)} = \frac{m}{n} + \frac{1}{n} \frac{\sin\left(\frac{\pi m}{n}\right)}{\sin\left(\frac{\pi }{n}\right)}\;.
\end{equation}
Using the inequality $\sin(m\pi/n)\leq m \sin(\pi/n)$, the maximum efficiency is at most $2 m/n$, i.e.~twice the fraction of nodes that can be individually addressed. In particular, the upper bound $ 2 m /n$ is reached when $n\gg m$ and $n\gg 1$.

\subsubsection{Cycle graphs}
Let us now consider a generic {\bf cycle graph} $C_n$. The QFI evaluates to
\begin{equation}
\mathcal{Q}_{C_n}(\gamma)=16t^2\, \text{Var}\,\left[\cos\left(\frac{2\pi k \mathbb{X}}{n}\right)\right]\;,
\end{equation}
where $\mathbb X$ is a random variable, with sample space $j \in \{0, \dots , n -1\}$ and corresponding
probabilities $\text{Pr}(j) = |\alpha_j|^2$.
Making use of Popoviciu's inequality, see Eq.~\eqref{popineq}, one can maximize the QFI over the initial preparation, which gives: 
 \begin{equation}
  \underset{\ket{\psi_0}}{\text{max}}\; \mathcal{Q}_{C_n}(\gamma)=
 \begin{cases}
	 16t^2  \qquad \qquad&\text{if $n$ even}\\
	 4t^2\left[1+\cos\left(\frac{\pi}{n}\right)\right]^2 \quad &\text{if $n$ odd}
	\end{cases}\,.
 \end{equation}
 The optimal preparation is a balanced superposition of the ground state $\ket{\xi_0}$ and highest excited state $\ket{\xi_{n/2}}$, where $\xi_0 = 2(1-\gamma)$, $\xi_{n/2}=2(1+\gamma)$, and the corresponding eigenvectors are defined in Eq.~\eqref{eigenvectorsCycle}. After a time $t$, it gives rise to the state
\begin{equation}\label{stateopt}
  \ket{\psi_t} = \frac{1}{\sqrt 2} e^{-i\xi_0 t}\ket{\xi_0} + \frac{1}{\sqrt 2} e^{-i\xi_{n/2} t}\ket{\xi_{n/2}} = \frac{1}{\sqrt{2n}}\, e^{-2i(\epsilon-\gamma)t}\,
  \scriptstyle\begin{pmatrix}
  \scriptstyle 1+e^{-4i\gamma t}\\
  \scriptstyle 1-e^{-4i\gamma t}\\
  \scriptstyle \vdots\\
  \scriptstyle 1+e^{-4i\gamma t}\\
  \scriptstyle 1-e^{-4i\gamma t}\\
  \end{pmatrix}\;.
\end{equation}

\begin{figure}
\begin{center}
\includegraphics[width = \textwidth]{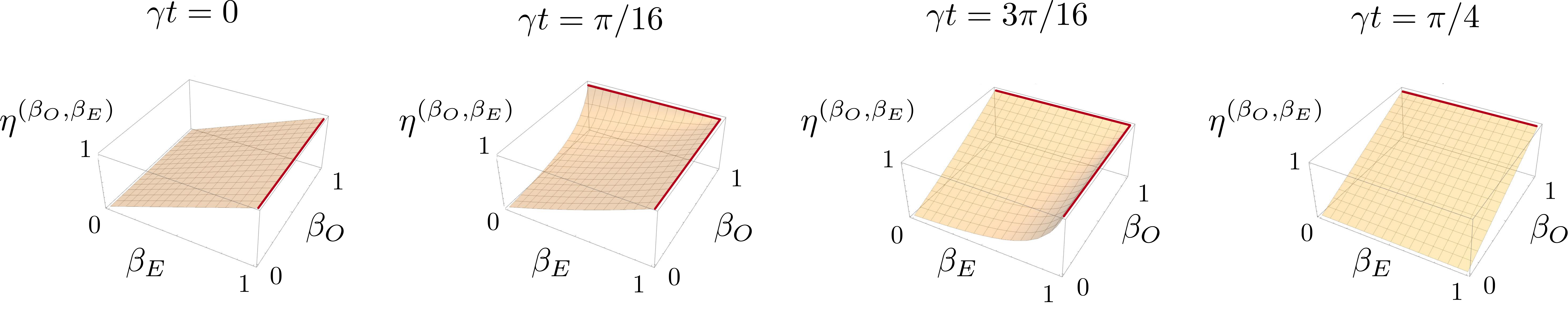}
\caption{The efficiency $\eta^{(\beta_O,\beta_E)}$ of an incomplete measurement, as a function of the fractions of odd nodes $\beta_O$ and even nodes $\beta_E$ under experimental control, for different values of $\gamma t$. Highlighted by a thick line (red), the optimal region of unit efficiency. For $\gamma t \neq m\pi/4$, $m\in \mathbb Z$, the optimal region consists of the two segments $(\beta_O,1)$ and $(1,\beta_E)$. For even multiples of $\pi/4$, only the first segment is present, while for odd multiples only the second.\label{FICycleFig}} 
\end{center}
\end{figure}

We now analyze the performance of incomplete position measurements taken on the state of Eq.~\eqref{stateopt}. The probability $p_j$ of measuring the walker at node $j$ is
\begin{equation}
  p_j = \frac{1}{n}\,[1+ (-1)^{j+1}\cos(4\gamma t)]= \begin{cases}
    \frac{2}{n} \cos^2(2\gamma t)\defeq p_O\qquad \text{if $j$ is odd}\\[5pt]
    \frac{2}{n} \sin^2(2\gamma t) \defeq p_E \qquad\, \text{if $j$ is even}
  \end{cases}\;.
\end{equation}
Let us assume that the experimentalist has access to  a subset of the graph's nodes, of which $n_O$ have an odd label and $n_E$ an even label. In the following, we take the total number of vertices $n$ to be even, without loss of generality. Introducing the notation $\beta_O\defeq 2n_O/n$ and $\beta_E\defeq 2n_E/n$ for, respectively, the fractions of odd and even nodes under individual control, the corresponding Fisher information $\mathcal F_{C_n}^{(\beta_0,\beta_E)}(\gamma)$ can be written as
\begin{equation}
   \mathcal F_{C_n}^{(\beta_0,\beta_E)}(\gamma) = n_O \frac{(\partial_\gamma p_O)^2}{p_O}+ n_E \frac{(\partial_\gamma p_E)^2}{p_E} + \frac{(\partial_\gamma \bar p)^2}{\bar p}\;,
\end{equation}
where $\bar p = 1- n_O p_O - n_E p_E$. After a standard computation, one finds:
\begin{equation}\label{FICycleDef}
  \mathcal F_{C_n}^{(\beta_0,\beta_E)}(\gamma) = 16 t^2\; \left[\frac{\beta_O + \beta_E-2 \beta_O\beta_E +(\beta_E-\beta_O)\cos(4\gamma t)}{2-(\beta_O+\beta_E)+(\beta_E-\beta_O)\cos(4\gamma t)}\right]\;.
\end{equation}
In spite of appearances, the previous expression is invariant under relabellings of the graph's nodes. A relabelling may change the parity of each vertex label, exchanging $\beta_O$ with $\beta_E$, and $p_O$ with $p_E$. It is now enough to make use of the relation $\cos(4\gamma t) = n p_O - 1 = 1- n p_E$  to check the invariance. From Eq.~\eqref{FICycleDef}, the efficiency of an incomplete measurement, optimized over the interrogation time $t$, has the following simple expression, i.e.,
\begin{equation}
\underset{t}{\text{max}}\;\eta^{(\beta_O,\,\beta_E)} = \text{max}(\beta_0,\,\beta_E)\;.
\end{equation}
It follows in particular that a complete position measurement is always optimal. Incomplete measurements can also be optimal, e.g.~a measurement of only the odd or only the even vertices still has unit efficiency (see also Fig.~\ref{FICycleFig}).
\subsection{Hypercube graphs}
An \emph{hypercube graph} $Y_d$ has eigenvalues $\xi_j$, where $j\in \{0,\dots,d\}$, see Eq.~\eqref{eigValuesHyper}. Each eigenvalue $\xi_j$ has multiplicity $[d,j]\defeq d!/j!(d-j)!$. The corresponding eigenstates $\ket{\xi_j^{(\kappa_j)}}$, with $\kappa_j\in\{1,\dots,[d,j]\}$, are constructed recursively for any $d$ by means of Eq.~\eqref{FRec}. The most general initial preparation is
	\begin{equation}
		\ket{\psi_0} = \sum_{j=0}^d \sum_{\kappa_j=1}^{[d,j]}\, \alpha_j^{(\kappa_j)} \ket{\xi_j^{(\kappa_j)}}\;,\qquad  \text{with} \quad \sum_{j=0}^d \sum_{\kappa_j=1}^{[d,j]} |\alpha_j^{(\kappa_j)}|^2 = 1\;.
	\end{equation}
 For future convenience, let us denote by $p_{\xi_j} \defeq \sum_{\kappa_j=1}^{[d,j]} |\alpha_j^{(\kappa_j)}|^2$ the total probability that an energy measurement returns the outcome $\xi_j$. It can be checked that the QFI at time $t$ and for a generic initial state can be written as:
 \begin{equation}
   \mathcal Q_{Y_d}(\gamma) = 4 t^2\, \text{Var} (d- 2\mathbb X)\;,
 \end{equation}
 where $\mathbb X$ is a random variable such that $\text{Pr}(j) = p_{\xi_j}$, for $j\in\{0,\dots,d\}$.
 Since the maximum value of $d- 2\mathbb X$ is equal to $d$ (when $\mathbb X =0$) and the minimum value is $-d$ (when $\mathbb X =d$), one has that  $\mathcal F_Q(\gamma) \leq 4 t^2 d^2$ by Popoviciu's inequality. The optimal QFI thus evaluates to  
 \begin{equation}
   \underset{\ket{\psi_0}}{\text{max}}\;\mathcal Q_{Y_d}(\gamma)=4 t^2 d^2,
 \end{equation}
  which scales quadratically with the dimension $d$ and the interrogation time $t$. The optimal preparation is a balanced superposition of the ground state $\ket{\xi_0}$ and the maximally excited state $\ket{\xi_d}$ (see Eq.~\eqref{groundexca}).

Let us now consider the performance of incomplete position measurements. After a time $t$, an incomplete position measurement is performed on the state $\ket{\psi_t} =  (e^{-i \xi_0 t} \ket{\xi_0}+ e^{-i \xi_d t} \ket{\xi_d})/\sqrt 2$, where $\xi_0 = d(1-\gamma)$, $\xi_d = d(1 + \gamma)$. We adopt the following notation: $\mathcal F_{Y_d}^{(\delta)}(\gamma)$ denotes the Fisher information for an incomplete measurement having as POVM the $2^\delta$ rank-1 projectors over the nodes
making up a $\delta$-dimensional face of the hypercube, plus the projector onto their orthogonal complement. It can be computed as
\begin{equation}\label{FIHyper}
   F_{Y_d}^{(\delta)}(\gamma) = 2^{\delta -1}\, \frac{(\partial_\gamma p_+)^2}{p_+}+ 2^{\delta -1} \frac{(\partial_\gamma p_-)^2}{p_-}\;,
\end{equation}
where
\begin{equation}
  p_+ \defeq \frac{1}{2^{d-1}}\,\cos^2(d\gamma t)\;,\qquad\quad p_- \defeq \frac{1}{2^{d-1}}\,\sin^2(d\gamma t)\;.
\end{equation}
Notice that the probability of finding the walker in any of the accessible 
nodes is $2^{\delta-1}p_+ + 2^{\delta-1}p_-=1/2^{d-\delta}$, which is,
 in particular, independent of $\gamma$. Thus, no term analogous to 
 the last one on the right-hand-side of Eq.~\eqref{FICompl} appears
  in Eq.~\eqref{FIHyper}. The FI evaluates to
\begin{equation}
   \mathcal F_{Y_d}^{(\delta)}(\gamma) = 2^{\delta - d + 2} d^2 t^2\;.
\end{equation}
Its efficiency is $\eta^{(\delta)} = 1/2^{d-\delta}$,
i.e.~the ratio between the number of nodes under individual control and the total number of nodes. It follows that, in particular, a complete measurement (when $\delta=d$) is  optimal.
\subsection{Complete bipartite graphs}
The generic complete bipartite graph $K_{p,q}$ has $n=p+q$ eigenvectors, of which only two, $\ket{\xi_{\pm}}$ given in Eq.~\eqref{xiDet}, depend on the parameter $\gamma$. All other eigenvectors, as well as their eigenvalues, are independent of $\gamma$; thus, no estimation strategy can fruitfully make use of them. As a consequence, the initial preparation is taken to be a superposition
 of $\ket{\xi_\pm}$ only, e.g.~$\ket{\psi_0} = \alpha_- \ket{\xi_-} +\alpha_+
 \ket{\xi_+}$. The corresponding QFI at the generic time $t$ evaluates to
 \begin{equation}\label{QFICBip}
   \mathcal Q_{K_{p,q}}(\gamma) = \frac{4\left(f_{p,q}-4 g_{p,q}^2\right)}{\Delta_{p,q}^2}\;,
 \end{equation}
 where
 \begin{align}
  & f_{p,q} = pq[16p^2 q^2 \gamma^4 t^2 + (p-q)^2 (1+4pq\gamma^2 t^2)]\;,\\
  & g_{p,q} = (\abs{\alpha_-}^2-\abs{\alpha_+}^2)\, pq \gamma t \sqrt{\Delta_{p,q}} +\Im\left(e^{it\sqrt{\Delta_{p,q}}} \bar \alpha_+ \alpha_-\right)(p-q)\sqrt{pq}\;.
 \end{align}
 The optimal initial preparation is such that $g_{p,q}$ vanishes, which 
 is obtained when $\abs{\alpha_-}=\abs{\alpha_+}$, with a relative phase 
 $\phi_{\text{opt}}=\text{arg}(\alpha_+/\alpha_-)=t \sqrt{\Delta_{p,q}}$. The maximum QFI is 
 therefore equal to 
 \begin{equation}
    \underset{\ket{\psi_0}}{\text{max}}\; \mathcal Q_{K_{p,q}}(\gamma)=4f_{p,q}/\Delta_{p,q}^2.
 \end{equation}
  For fixed number of vertices $n$, one may further optimize over the cardinality of each bipartition $p$ and $q$. The maximum is reached for $p = \lfloor n/2 \rfloor$, which leads to a corresponding scaling  $\sim n^2 t^2$, quadratic both in the number of vertices and the interrogation time.
\par
 We now study in some detail the special case of a {\bf star graph} $S_n$. The maximum QFI is obtained via the substitutions $p=1$ and $q=n-1$,
 \begin{equation}
\underset{\ket{\psi_0}}{\text{max}}\; \mathcal Q_{S_n}(\gamma) =  \frac{4(n-1)[16(n-1)^2\,\gamma^4 t^2+(n-2)^2[1+4(n-1)\gamma^2 t^2]]}{[(n-2)^2 + 4 (n-1)\gamma^2]^2}\;.
 \end{equation}
It depends on $n$, but does not grow indefinitely with the size of the graph. When $n\to \infty$, it saturates instead to a constant value. Therefore, an optimal number of nodes $n_{\text{opt}}$ may exist. We solve for $n_{\text{opt}}$ in the two opposite regimes of small and long times. For small times $\gamma t \ll 1$,
 \begin{equation}
\underset{\ket{\psi_0}}{\text{max}}\; \mathcal Q_{S_n}(\gamma) \sim \frac{4(n-1)(n-2)^2}{[(n-2)^2+4(n-1)\gamma^2]^2}\;,
 \end{equation}
 which is maximized by
 \begin{equation}
   n_{\text{opt}} \sim 2[1+\gamma^2 +\gamma\sqrt{1+\gamma}].
 \end{equation}
For large times $\gamma t \gg 1$,
 \begin{equation}
\underset{\ket{\psi_0}}{\text{max}}\; \mathcal Q_{S_n}(\gamma) \sim \frac{16\gamma^2 t^2(n-1)[4(n-1)^2\gamma^2+(n-2)^2(n-1)]}{[(n-2)^2  + 4(n-1)\gamma^2]^2}\;.
 \end{equation}
 If $1/\gamma^2 \leq 2$, then there is no optimal value of $n$ (the optimal value is $n=\infty$). Instead, if $1/\gamma^2 > 2$, the optimal value of $n$ is
 \begin{equation}
   n_{\text{opt}} \sim \frac{2(1/\gamma^2-1)}{(1/\gamma^2-2)}\;.
 \end{equation}
 Adding new vertices above $n_{\text{opt}}$ will lower the maximum achievable precision.

\begin{figure}
\centering
\includegraphics[width=0.8\textwidth]{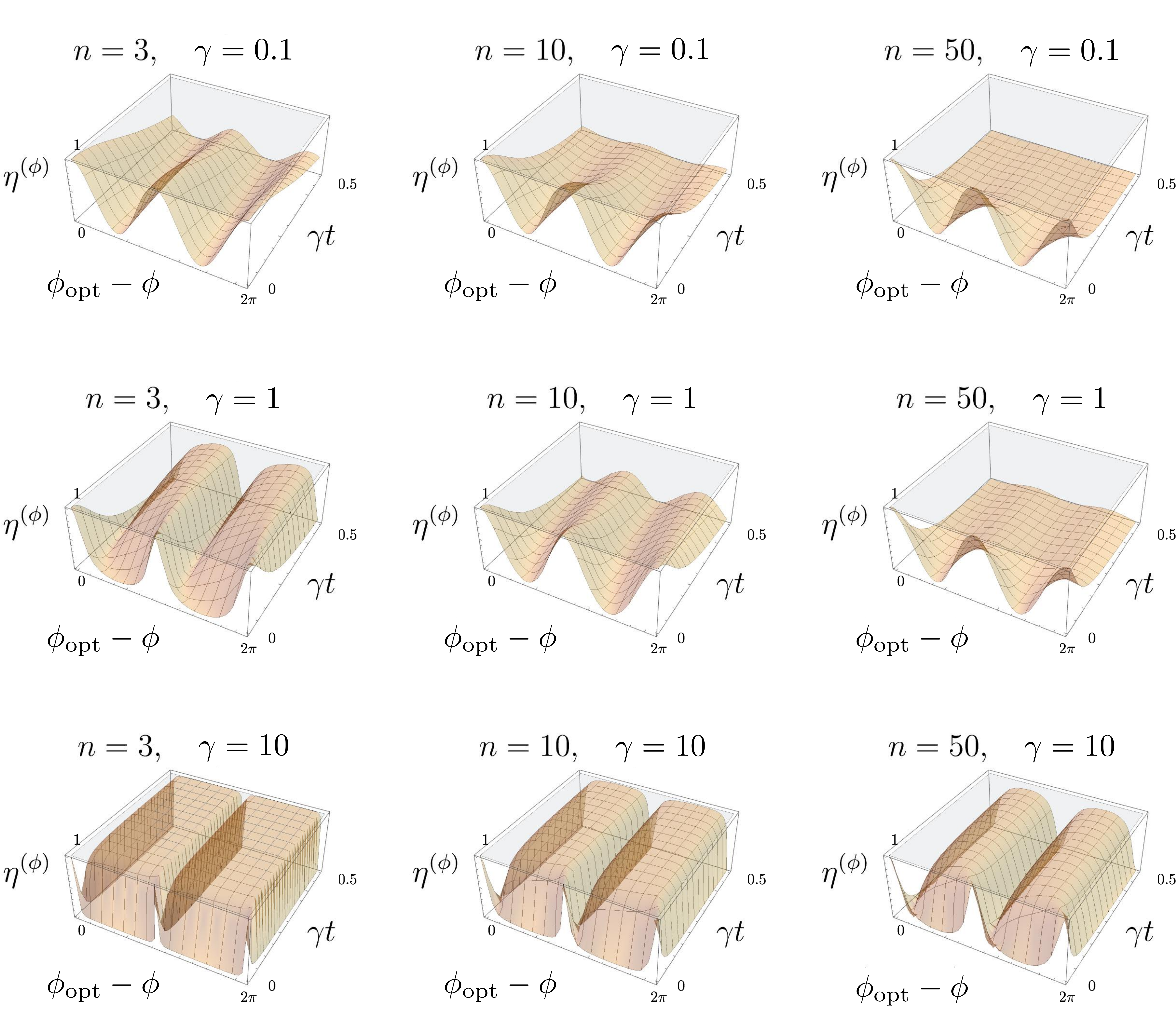}
\caption{The efficiency $\eta^{(\phi)}$ of a complete measurement, as a function of the phase difference $\phi_{{\text{opt}}}-\phi$ and of the dimensionless time scale $\gamma t$ (for different values of $n$ and $\gamma$). In general, the efficiency is closer to one for higher values of $\gamma$ and for smaller values of $\gamma t$ and of $n$. Only for $t=0$, there always exist a choice of $\phi$ which allows to reach unit efficiency; otherwise, a complete measurement is suboptimal.\label{StarEff}}
\end{figure}

From our previous discussion, the optimal initial preparation is a balanced superposition of the two energy eigenstates $\ket{\xi_\pm}$ that can be read off from Eq.~\eqref{eigvecspm} after setting $p=1$ and $q=n-1$, with a relative phase $\phi_{\text{opt}}= t \sqrt{\Delta_{1,n-1}}$. Since the optimal phase $\phi_{\text{opt}}$ depends on $\gamma$, an adaptive procedure is required in order to extract the maximum QFI. For the moment, we assume that the walker is prepared in the state
$\ket{\psi_0} = (\ket{\xi_-} + e^{i\phi} \ket{\xi_+})/\sqrt 2$, where $\phi\in [0,2\pi]$ is arbitrary. Let us suppose that at time $t$ an incomplete position measurement is performed. First, we consider the case of an incomplete measurement monitoring only the central node, with associated POVM made up of the two projectors $\ket{1}\bra{1}$ and $\mathbb I_n- \ket{1}\bra{1}$. One finds that its Fisher information coincides with the FI for a complete position measurement, denoted by $\mathcal F_{S_n}^{(\phi)}(\gamma)$ (the superscript makes manifest the dependence on the arbitrary phase $\phi$ of the initial state). The implication is that distinguishing outcomes corresponding to the walker being in one peripheral node or the other is useless for estimation purposes: one may as well monitor only the central node. The efficiency $\eta^{(\phi)}$ of a position measurement (either a complete measurement or an incomplete one, but including the central node) is
\begin{equation}
  \eta^{(\phi)} =\; \frac{(n-1)[(n-2)^2  \cos(\phi_{\text{opt}}-\phi)-4(n-1)\sqrt{\Delta_{1,n-1}}\, \gamma^2 t \sin(\phi_{\text{opt}}-\phi)]^2}{f_{1,n-1}\left[4(n-1)\gamma^2 \sin^2(\phi_{\text{opt}}-\phi)+(n-2)^2\right]}\;.
\end{equation}
Except for	a few special cases (when $n=2$, or $t=0$ and $\phi = \phi_{\text{opt}}$), a position measurement is always  suboptimal. For short interrogation times, expanding for $\gamma t \ll 1$ and  $\phi = \phi_{\text{opt}}$, one obtains
\begin{equation}
  \eta^{(\phi_{\text{opt}})} = 1 - \frac{4(n-1) \Delta_{1,n-1}}{(n-2)^4}\gamma^2t^2 + o(\gamma^2 t^2)\;,\qquad \qquad (\gamma t \to 0)\;.
\end{equation}
For large number of vertices $n\to \infty$ and $\phi = \phi_{\text{opt}}$, one has instead
\begin{equation}
  \eta^{(\phi_{\text{opt}})} = \frac{1}{4\gamma^2 t^2 n}+o(1/n)\;,\qquad \qquad (n\to \infty),
\end{equation}
i.e.~the efficiency decreases linearly with the number of vertices. The reader is also referred to Fig.~\ref{StarEff} for more details about the different possible regimes.

\begin{table}[h!]
\resizebox{\columnwidth}{!}
{\begin{tabular}{!{\vrule width 1pt}c|c|c|c|c!{\vrule width 1pt}}
\thickhline
  & \textbf{scaling of QFI with $n$} & \textbf{optimal preparation} & \textbf{optimal measurement} & \textbf{$\eta$ of incomplete position measurement} \TBstrut \\ \thickhline
  $K_n$ &  $\sim n^2$ & \makecell{any balanced superposition of ground\\ state and any other excited state} & position & $\beta + 1/n\cdot\sin(\pi\beta)/ \sin(\pi/n)$ \TBstrut \\ \hline
  $C_n$ & \makecell{$\begin{cases} $independent of $ n &\mbox{if $n$ even } \\
\sim \left[1+\cos\left(\frac{\pi}{n}\right)\right]^2 & \mbox{if $n$ odd}  \end{cases}$} 
  & \makecell{any balanced superposition of ground\\ state and highest excited state} & position & $\text{max}(\beta_O,\,\beta_E)$ \TBstrut\\ \hline
  $Y_d$ &  $\sim (\log n)^2$ & \makecell{any balanced superposition of ground \\ state and highest excited state} & position & $\beta$ \\ \hline
 $S_n$ &  $\exists\; n_{\text{opt}}$ &  \makecell{balanced superposition of $\ket{\xi_{\pm}}$\\ with relative phase $\phi_{\text{opt}}$} &  exotic   & independent of $\beta$\TBstrut\\ \thickhline
\end{tabular}}
\caption{For each family of graphs considered in the main text, we report the scaling of the QFI with the total number of nodes $n$, the optimal measurement saturating the quantum Cram\'er-Rao bound, the optimal initial preparation and the efficiency $\eta$ of an incomplete position measurement. Notice that $\beta$ (\emph{resp}., $\beta_O$, $\beta_E$) denotes the fraction of the graphs's nodes (\emph{resp}., nodes with even labels, odd labels) under individual control by the experimentalist.}\label{tabQW}
\end{table}
\section{Conclusions} \label{sec:conc}
In this paper, we have studied the problem of estimating the tunnelling amplitude  $\gamma$  for a quantum walker evolving continuously in 
time on a graph $G$, where $G$ is an element of a few relevant 
families of graphs. Our first result is that the topology of the 
graph may have dramatic effects on the maximum extractable information. 
For instance, the quantum Fisher information exhibits  different scaling laws with the total number of vertices (see Table \ref{tabQW}). For each family considered, we have maximized the quantum Fisher information over the initial preparation, determining the optimal initial state of the walker, as well as the optimal measurement. 
\par
We have then discussed in details the performance of position 
measurements. Complete position measurements, which may implemented
when one has experimental access to the full set of graph's node,
perform quite well: they are often optimal, the only exception being 
(among the cases taken into consideration) that of complete bipartite graphs, e.g.~star graphs. Incomplete, e.g. nearly-local, position measurements still allow to extract a non-vanishing amount of information. Their efficiency (i.e.~the ratio of the FI to QFI) is closely related to the fraction $\beta$ of the graph's nodes that are under control by the experimentalist. The exception is again the case of star graphs, since  monitoring only the central node yields the same information as monitoring each node separately.
\par
Our results uncover fundamental properties of quantum walks 
related to their topologies, and pave the way to optimal 
design of quantum walks implementation, e.g. with superconducting 
circuits.
\section*{Acknowledgements}
MGAP is member of GNFM-INdAM.
\appendix
\section{Spectral properties of selected graphs}\label{appe}
\setcounter{section}{1}
\subsection*{Circulant graphs: complete  and cycle graphs}
The adjacency matrix of a circulant graph $O_n$ is a circulant matrix, i.e.~each row is obtained by shifting the preceding one to the right. Denoting by $d$ the degree of each vertex, we consider the following quantum walk Hamiltonian on $O_n$:
\begin{equation}\label{circulantHMat}
H_{O_n} =\begin{pmatrix}
d  & -\gamma_1 & -\gamma_2  & \dots & -\gamma_{n-1}\\
-\gamma_{n-1} & d & -\gamma_1 & \dots & -\gamma_{n-2}\\
-\gamma_{n-2} & -\gamma_{n-1} & d& \dots & -\gamma_{n-3}\\
\vdots & \vdots & \vdots & \ddots & \vdots\\
-\gamma_1 & -\gamma_2 & -\gamma_3 & \dots & d
\end{pmatrix}\;.
\end{equation}
We impose the symmetry constraint $\gamma_j = \gamma_{n-j}$, so that $H_{O_n}$ is itself a circulant matrix. Notice that there are a total of $\lfloor n/2 \rfloor$ independent couplings. In particular, if all weights are set equal to the same value, Eq.~\eqref{circulantHMat} reduces to the case of a complete graph $K_n$; if instead the only non-zero weights are $\gamma_1=\gamma_{n-1}$ one has a cycle graph $C_n$.

From the general theory of circulant matrices \cite{aldrovandi_special_2001}, the eigenvalues of $H_{O_n}$ are found to be
\begin{equation}\label{geneigval}
\xi_j= d - \sum_{k=1}^{n-1} \gamma_k\, e^{\frac{2\pi i jk}{n}}\;,\qquad\qquad j\in\{0,\dots, n-1\}\;.
\end{equation}
Using the fact that $\gamma_j = \gamma_{n-j}$, one may rewrite the previous equation as
\begin{equation*}
  \xi_j =\begin{cases}
 d - 2 \gamma_1 \cos\left(\frac{2\pi j}{n}\right)  -\dots - 2\gamma_{(n-2)/2}  \cos\left(\frac{(n-2)\pi j}{n}\right) - \gamma_{n/2} \cos\left(\pi j\right)\quad\, \text{($n$ even)}\\[10pt]
 d  - 2 \gamma_1 \cos\left(\frac{2\pi j}{n}\right)  - \dots - 2 \gamma_{(n-1)/2} \cos\left(\frac{(n-1)\pi j}{n}\right)\quad\qquad \qquad \;\, \qquad  \text{($n$ odd)}
  \end{cases}\;.
\end{equation*}
Notice that the spectrum is doubly degenerate, i.e.~$\xi_{j} = \xi_{n-j}$. The eigenvectors are
\begin{equation}
\label{eigvec}
\ket{\xi_j} = \frac{1}{\sqrt n}\sum_{k=1}^{n} e^{\frac{2\pi i j (k-1)}{n}}\,\ket{k}\;.
\end{equation}
\subsubsection*{Complete graphs}
	When all couplings are equal among themselves, the Hamiltonian of Eq.~\eqref{circulantHMat} reduces to:
	\begin{equation}\label{HComplete}
	H_{K_n} =\begin{pmatrix}
	(n-1)  & -\gamma & -\gamma  & \dots & -\gamma \\
	-\gamma  & (n-1)  & -\gamma & \dots & -\gamma \\
	\vdots & \vdots & \vdots & \ddots & \vdots\\
	-\gamma  & -\gamma  & -\gamma  & \dots & (n-1)
	\end{pmatrix}\;.
	\end{equation}
 	Making use of Eq.~\eqref{geneigval} and the identity $ \sum_{k=0}^{n-1} \text{exp}(2\pi jk/n)=0$, the eigenvalues of $H_{K_n}$ can be written more compactly as
	\begin{equation}\label{eigvaluesCompletea}
	\xi_j  =
	\begin{cases}
	 (n-1)(1 - \gamma)  \qquad \text{if $j=0$}\\
	 (n-1) + \gamma \;\quad\qquad \text{if $j\neq 0$}
	\end{cases}\;.
	\end{equation}
	The least eigenvalue is $\xi_0$, whereas the remaining eigenvalues are all degenerate. The corresponding eigenvectors are the same as in Eq.~\eqref{eigvec}.
\subsubsection*{Cycle graphs}
	For cycle graphs, all couplings are zero except for $\gamma_1 = \gamma_{n-1}$.
	The Hamiltonian $H_{C_n}$ is
	\begin{equation}\label{HMatCycle}
	H_{C_n} =\begin{pmatrix}
	2  & -\gamma & 0  & \dots & -\gamma \\
	-\gamma  & 2  & -\gamma & \dots & 0 \\
	0 & -\gamma  & 2 & \dots & 0 \\
	\vdots & \vdots & \vdots & \ddots & \vdots\\
	-\gamma  & 0  & 0  & \dots & 2
	\end{pmatrix}\;.
	\end{equation}
	In the main text,  the case of a cycle graph with an even number of vertices is considered. Specializing some of the above formulas, the least eigenvalue of $H_{C_n}$ is found to be $\xi_0=2(1-\gamma)$, while the largest is $\xi_{n/2}=2(1+\gamma)$, with corresponding eigenvectors
	\begin{equation}\label{eigenvectorsCycle}
\ket{\xi_0} = \frac{1}{\sqrt n} (1,1\dots,1,1)^t  \qquad \text{and} \qquad
	\ket{\xi_{n/2}} = \frac{1}{\sqrt n} (1,-1\dots,1,-1)^t \;.
\end{equation}

\subsection*{Hypercube graphs}
For a generic hypercube graph $Y_d$, its quantum walk Hamiltonian can be written as $H_{Y_d} = d\, \mathbb I_{2^d} -\gamma A^{(d)}$, where $A^{(d)}$ is 
the adjacency matrix of $Y_d$, which is defined recursively via the following relation:
\begin{equation}
		A^{(1)} = \begin{pmatrix}
    0 & 1 \\
    1 & 0 \\
  \end{pmatrix}\;, \qquad \qquad
	A^{(d)} = \begin{pmatrix}
   A^{(d-1)} & \mathbb I_{2^{d-1}} \\
   \mathbb I_{2^{d-1}} & A^{(d-1)}
  \end{pmatrix}\;.
\end{equation}
The eigenvectors of $H_{Y_d}$ are denoted by $\ket{\xi_j^{(\kappa_j)}}$, where $j\in \{0,\dots,d\}$ and $\kappa_j$ is a degeneracy index ranging from 1 to $[d,j] \defeq d!/j!(d-j)!$. The corresponding eigenvalues are $\xi_j = d - \gamma(d-2j)$. The eigenvectors $\ket{\xi_j^{(\kappa_j)}}$ coincide with the columns of a sequence of matrices $B_d$, indexed by the dimension $d$ and defined recursively as follows:
\begin{equation}\label{FReca}
B_1 = \frac{1}{\sqrt 2} \begin{pmatrix}
1 & 1 \\
1 & -1
\end{pmatrix}\;,\qquad \qquad
B_d = \frac{1}{\sqrt 2}\begin{pmatrix}
B_{d-1} & B_{d-1} \\
B_{d-1} & - B_{d-1}
\end{pmatrix}\;.
\end{equation}
By construction, each $B_d$ is a Hadamard matrix. In particular, the ground state $\ket{\xi_0}$ 
and the highest excited state $\ket{\xi_d}$ can be written as:
\begin{equation}\label{groundexca}
\ket{\xi_0}  = \frac{1}{2^{d/2}}\begin{pmatrix}
  1\\
  1
  \end{pmatrix}^{\!\otimes\, d}\;,\qquad \qquad \ket{\xi_d}  = \frac{1}{2^{d/2}} \begin{pmatrix}
  1\\
  -1
  \end{pmatrix}^{\!\otimes\, d}\;.
\end{equation}
Notice that they are non-degenerate, i.e.~$\kappa_0=\kappa_d=1$, so we omit the degeneracy label.
%
\subsection*{Complete bipartite graphs}
For a complete bipartite graph $K_{p,q}$, the quantum walk Hamiltonian $H_{K_{p,q}}$ takes the following block form:
\begin{equation}
  H_{K_{p,q}} = \left(\begin{array}{c|c}
  q\, \mathbb I_p & -\gamma \mathbb J_{p\times q} \\
  \hline
  -\gamma \mathbb J_{q\times p} & p\, \mathbb I_q
  \end{array}\right)\;,
\end{equation}
where $\mathbb I_p$ is the $p\times p$ identity matrix and $\mathbb J_{p\times q}$ is the $p\times q$ matrix made up of all ones. In the position eigenbasis, a natural ansatz for the generic eigenvector $\ket{\xi}$ of $H_{K_{p,q}}$ is in the form $\ket{\xi} =(\boldsymbol x,\,\boldsymbol y)^t$, where $\boldsymbol x$ is a $p$-vector and $\boldsymbol y$ a $q$-vector. The eigenvalue equation $H_{K_{p,q}} \ket{\xi} = \xi\cdot \ket{\xi}$
 implies the linear system of constraints:
\begin{equation}\label{sysCB}
  \begin{cases}
    (q -\xi ) \boldsymbol x = \gamma \mathbb J_{p\times q}\, \boldsymbol y\\
    (p -\xi) \boldsymbol y = \gamma \mathbb J_{q \times p}\, \boldsymbol x
  \end{cases}\;.
\end{equation}
Multiplying by $\mathbb J_{q\times p}$ the first of the previous equations and by $\mathbb J_{p\times q}$ the second equation, and using the fact that $\mathbb J_{q\times p}\, \mathbb J_{p\times q} = p\, \mathbb J_{q\times q}$, one obtains:
\begin{equation}\label{c1CB}
  \mathbb J_{q\times q}\, \boldsymbol y = \frac{(q -\xi)(p -\xi)}{\gamma^2 p}\,\boldsymbol y\;, \qquad
    \mathbb J_{p\times p}\, \boldsymbol x = \frac{(q -\xi)(p -\xi)}{\gamma^2 q}\,\boldsymbol x\;.
\end{equation}
It follows that $\boldsymbol y$ is an eigenvector of $\mathbb J_{q\times q}$and  $\boldsymbol x$ is an eigenvector of $\mathbb J_{p\times p}$. Let us recall that the spectrum of $\mathbb J_{p\times p}$ is made up of the eigenvalue $0$ (with multiplicity $p-1$ and corresponding eigenspace spanned by all $p$-vectors whose components sum to zero) and the eigenvalue $p$ (with multiplicity 1 and corresponding eigenvector $\boldsymbol 1_p$, the $p$-vector made up of ones). Eqs.~\eqref{c1CB} thus imply that the only possible values for $\xi$ are $\text{spec}(H_G) =
\{\xi_1,\,\xi_2,\,\xi_+,\xi_-\}$, where $\xi_1 \defeq q$, $\xi_2 \defeq p$ and $\xi_\pm$ are the roots of the equation $(q- \xi_\pm)(p-\xi_\pm) = qp\, \gamma^2$, i.e.
\begin{equation}
  \xi_{\pm} = \frac{(p+q)\pm\sqrt{\Delta_{p,q}}}{2}\;,\qquad \quad \text{with}\qquad \Delta_{p,q} \defeq (p-q)^2 +4qp\,\gamma^2\;.
\end{equation}
Notice that the lowest eigenvalue is $\xi_-$, the highest is $\xi_+$, while $\xi_1$ and $\xi_2$ are always in between.

The corresponding eigenvectors can be found as follows. For the eigenvalue $\xi_1$, one finds that $\boldsymbol y$ must vanish and that $\boldsymbol x \in \text{ker}\,{\mathbb J_{p\times p}}$, whereas for $\xi_2$, $\boldsymbol x$ must vanish and $\boldsymbol y \in \text{ker}\,{\mathbb J_{q\times q}}$. Introducing two orthonormal basis $\boldsymbol a^{(\kappa)}$ and
$\boldsymbol b^{(\kappa')}$, for $\text{ker}\,{\mathbb J_{p\times p}}$ and $\text{ker}\,{\mathbb J_{q\times q}}$ respectively, one can write:
\begin{equation}\label{eigvecs}
  \ket{\xi_1^{(\kappa)}} = (\boldsymbol a^{(\kappa)},\boldsymbol 0_q)^t\;,\qquad \qquad
  \ket{\xi_2^{(\kappa')}} = (\boldsymbol 0_p,\boldsymbol b^{(\kappa')})^t\;,
\end{equation}
where $\kappa\in \{1,\dots,p-1\}$ and $\kappa'\in \{1,\dots,q-1\}$. For the remaining two eigenvalues $\xi_\pm$, after substituting into the eigenvalue equation, one finally finds
\begin{equation}\label{eigvecspm}
  \ket{\xi_\pm} = \eta_\pm \begin{pmatrix}
\frac{p  - \xi_\pm}{\gamma p}\,\boldsymbol 1_p\\
  \boldsymbol 1_q
  \end{pmatrix}\;,\qquad \quad \text{with}\qquad  \eta_\pm \defeq \left[q\left(1+\frac{p-\xi_\pm}{q - \xi_\pm}\right) \right]^{-1/2}\;.
\end{equation}
Since Eqs.~\eqref{eigvecs} and \eqref{eigvecspm} already define a set of $p+q$ orthonormal eigenvectors, there are no additional eigenvectors.

\end{document}